\documentclass[aps,prd,amsmath,twocolumn,amssymbaps,showpacs]{revtex4-2}
\usepackage{graphicx}  
\usepackage{dcolumn}   
\usepackage{bm}        
\usepackage{amssymb}   
\usepackage{amsmath}   
\usepackage{bbm}
\usepackage{draftcopy}
\usepackage{relsize} 
\usepackage{xfrac}    
\usepackage{slashed}

\usepackage{color}
\definecolor{dgreen}{cmyk}{1.,0.,1.,0.2}        
\definecolor{orange}{cmyk}{0.,0.353,1.,0.}    

\usepackage[bookmarks]{hyperref}

\newcommand{\tr}{{\rm tr}}

\newcommand{\be}{\begin{equation}}
\newcommand{\ee}{\end{equation}}                                                                               
\newcommand{\bea}{\begin{eqnarray}}
\newcommand{\eea}{\end{eqnarray}}

\begin{document}

\title{Topological transition in a parallel electromagnetic field}

\author{Gaoqing Cao}
\affiliation{School of Physics and Astronomy, Sun Yat-sen University, Zhuhai 519088, China}
\date{\today}

\begin{abstract}
In this work, we attack the problem of "chiral phase instability" ($\chi$PI) in a quantum chromodynamics (QCD) system under a parallel and constant electromagnetic field. The $\chi$PI refers to that: When $I_2\equiv{\bf E\cdot B}$ is larger than the threshold $I_2^c$, no homogeneous solution can be found for $\sigma$ or $\pi^0$ condensate, and the chiral phase $\theta$ becomes unstable. Within the two-flavor chiral perturbation theory,  we obtain an effective Lagrangian density for $\theta(x)$ where the chiral anomalous Wess-Zumino-Witten term is found to play a role of "source" to the "potential field" $\theta(x)$. The Euler-Lagrangian equation is applied to derive the equation of motion for $\theta(x)$, and physical solutions are worked out for several shapes of system. In the case $I_2>I_2^c$, it is found that the $\chi$PI actually implies an inhomogeneous QCD phase with $\theta(x)$ spatially dependent. By its very nature, the homogeneous-inhomogeneous phase transition is of pure topological and second order at $I_2^c$. Finally, the work is extended to the three-flavor case, where an inhomogeneous $\eta$ condensation is also found to be developed for $I_2>I_2^c$. Correspondingly, there is a second critical point, $I_2^{c'}=24.3I_2^c$, across which the transition is also of topological and second order by its very nature.
\end{abstract}

\maketitle

\section{Introduction}
Recently, high energy nuclear physicists are devoting more energy to looking for the signals of chiral magnetic effect (CME) in experiments~\cite{Kharzeev:2022hqz,Chen:2023jhx,Xu:2023elq,Wang:2023yis,STAR:2023gzg} since it was denied by the blind analysis of the Bean Energy Scan II data from Relativistic Heavy Ion Collider~\cite{STAR:2021mii}. As well known, CME is the first proposed macroscopic manifestation of chiral anomaly~\cite{Adler:1969gk,Bell:1969ts} transport: In the presence of a finite chiral density, a constant magnetic field could induce a finite nondissipative electric current along its direction~\cite{Kharzeev:2007jp,Fukushima:2008xe}. After that, people discovered other anomalous transport phenomena that are also related to electromagnetic (EM) field, such as chiral magnetic wave~\cite{Kharzeev:2010gd}, chiral separation effect~\cite{Son:2004tq,Metlitski:2005pr}, chiral electric separation effect~\cite{Huang:2013iia}, anomalous magnetovorticity effect~\cite{Hattori:2016njk}, and chiral electric vortical effect~\cite{Cao:2021jjy,Yamamoto:2021gts}. Especially, only EM field and vorticity are involved in the last two effects, and the corresponding density-currents could be uniformly presented in Lorentz covariant forms~\cite{Cao:2021jjy}.

However, even restrict to the many-body effect, chiral anomaly could manifest itself not only through anomalous transports but also by modifying the ground states of the systems. In a system with finite baryon chemical potential and magnetic field, chiral separation effect is expected on one hand~\cite{Son:2004tq,Metlitski:2005pr}, but the ground state can be reorganized as $\pi^0$ domain wall~\cite{Son:2007ny} (or more precisely, chiral soliton lattice~\cite{Brauner:2016pko}) on the other hand. In a system with parallel and constant EM field, the chiral density would increase with time on one hand; but on the other hand, the ground state would rotate from the one with homogeneous $\sigma$ condensate to the one with both homogeneous $\sigma$ and $\pi^0$ condensates~\cite{Cao:2015cka,Wang:2017pje,Wang:2018gmj}. All the mentioned studies on the ground states were carried out within the chiral perturbation theory (ChPT)~\cite{Son:2007ny,Brauner:2016pko,Cao:2015cka}, a precise effective theory of QCD at low energy level, so the findings are reliable and model independent.

Considering the case with a parallel EM field further, our previous studies~\cite{Cao:2015cka,Cao:2019hku,Cao:2020pjq} had shown that homogeneous solutions could not be found when the second Lorentz invariant of EM field, $I_2\equiv{\bf E\cdot B}$, is large. Actually, if we check the thermodynamic potential for the chiral angle, extremal minimum that satisfies gap equation does not exist at all; while the boundary minimum strongly depends on the domain and does not satisfy the gap equation~\cite{Cao:2019hku}. In this sense, we interpreted that regime as a "chaotic" phase where the chiral angle is unstable and keeps changing with time. However, we are not satisfied with these arguments to address the problem of “chiral phase instability” ($\chi$PI), especially regarding that no static solution exists. Recall that there is no continuous symmetry breaking in two-dimensional thermal systems due to the instability caused by phase fluctuations~\cite{Coleman:1973ci,Mermin:1966fe,Hohenberg:1967zz}, then the Berezinskii-Kosterlitz-Thouless transition~\cite{Berezinsky:1970fr,Berezinsky:1972rfj,Kosterlitz:1973xp} is allowed to emerge as a completely new topological phase transition. So, it is natural to ask whether the $\chi$PI similarly implies a pure topological transition in a three-dimensional QCD system? If so, this could be the first time that a pure topological transition is found in a QCD system to the best of our knowledge.

The paper is devoted to addressing this problem and is organized as follows. We demonstrate and solve the problem of “chiral phase instability” in great detail within the two-flavor chiral perturbation theory (ChPT)  in Sec.\ref{2f}, where Sec.\ref{SS} and Sec.\ref{CS} are devoted to spherical and cylindrical systems, respectively. Take the spherical system for example, we show how the equation of motion can be solved for the chiral angle in chiral limit in Sec.\ref{CL} and in the real case in Sec.\ref{RC}. Moreover, Sec.\ref{nature} is set up to amply discuss the nature of phase transition across $I_2^c$. In Sec.\ref{3f}, the study is extended to the three-flavor case for a cylindrical system of infinitely long. Finally, we give a summary and discuss the prospects in Sec.\ref{sum}.

\section{Two-flavor chiral perturbation theory }\label{2f}
The Lagrangian of two-flavor chiral perturbation theory is given by~\cite{Gasser:1983yg}
\begin{eqnarray}
\label{chirall}
{\cal L}={\cal L}_0+{\cal L}_{\rm WZW},
\end{eqnarray}
where the normal chiral Lagrangian ${\cal L}_0$ is
\begin{eqnarray}
{\cal L}_0=\frac{f_\pi^2}{4}\tr\left[ D_\mu U^\dagger D^\mu U+m_\pi^2 (U+U^\dagger)\right]\label{L0}
\end{eqnarray}
up to $O(p^2)$ and ${\cal L}_{\rm WZW}$ is the Wess-Zumino-Witten (WZW) term given by~\cite{Wess:1971yu,Witten:1983tw}
\begin{eqnarray}
&&{\cal L}_{\rm WZW}=\frac{N_c}{48\pi^2}A_\mu \epsilon^{\mu\nu\alpha\beta}[\tr\left( QL_\nu L_\alpha L_\beta+Q R_\nu R_\alpha R_\beta\right)\nonumber\\&&\;\;\;\;\;\;\;\;\;\;\;\;\;\;\;\;\;\;\;\;\;\;\;\;\;\;\;\;\;\;\;\;\;\;\;\;\;\;\;-iF_{\alpha\beta} T_\nu],\label{LWZW}\\
&&T_\nu=\tr\left[ Q^2(L_\nu+R_\nu)+\frac{1}{2}\left( QU^\dagger QUL_\nu+QUQU^\dagger R_\nu\right)\right],\nonumber\\
&&L_\mu= \partial_\mu U^\dagger U ,\ \ \ R_\mu=U\partial_\mu U^\dagger.
\end{eqnarray}
In Eq.\eqref{L0}, the covariant derivative is defined as
\begin{eqnarray}
D_\mu U=\partial_\mu U+A_\mu[Q,U],
\end{eqnarray}
where $U$ is a $2\times2$ unitary matrix representing chiral fields in flavor space, $A_\mu$ is the vector potential for background EM field, and $Q={\rm diag}(2/3,-1/3)e$ is the charge matrix. By following the Weinberg parametrization, we have
\begin{eqnarray}
U=\frac{1}{f_\pi}(\sigma+i\boldsymbol{\tau\cdot \pi}),
\end{eqnarray}
where $\boldsymbol{\tau}$ is the Pauli matrix in flavor space, and the sigma boson $\sigma$ and pions $\boldsymbol{\pi}$ fulfill the constraint $\sigma^2+\boldsymbol{\pi}^2=f_\pi^2$. Note that the traces ${\rm tr}$ are all over flavor space in both Eqs.\eqref{L0} and \eqref{LWZW} and $\epsilon^{\mu\nu\alpha\beta}$ is the antisymmetric Levi-Civita symbol in space-time coordinates.

In the presence of a constant parallel electromagnetic field, we have already known that the phase with nonzero $\langle\sigma\rangle$ and $\langle\pi^0\rangle\equiv\langle\pi_3\rangle$ is favored~\cite{Cao:2015cka}. Here, we more generally assume that 
\bea
\langle\sigma\rangle=f_\pi \cos \theta(x),
\langle\pi^0\rangle=f_\pi \sin \theta(x)\label{sp}
\eea
with $\theta(x)$ the chiral phase or angle, and it follows from the constraint $\langle\sigma\rangle^2+\langle\boldsymbol{\pi}\rangle^2=f_\pi^2$ that $\langle\pi_1\rangle=\langle\pi_2\rangle=0$. Then, explicit calculations show that the normal Lagrangian density Eq.\eqref{L0} reduces to
\begin{eqnarray}
\bar{\cal L}_0&=&{f_\pi^2\over2}\partial_\mu\theta(x)\partial^\mu\theta(x)+f_\pi^2m_\pi^2\cos \theta(x).
\end{eqnarray}
To evaluate the explicit form of the WZW term, we find the expectation values (EVs) of ${L}_\mu$ and ${R}_\mu$ to be 
\bea
\bar{L}_\mu=\bar{R}_\mu=-i\,\tau_3\,\partial_\mu\theta(x),
\eea
which then implies the EV of $T_\nu$ to be
\bea
\bar{T}_\nu=3\,\tr\, Q^2\bar{L}_\nu=-i\,e^2\partial_\nu\theta(x).
\eea
Due to the commutability among $\bar{L}_\nu, \bar{L}_\alpha$ and $\bar{L}_\beta$, it is easy to check that the cubic terms of $\bar{L}_\mu$ and $\bar{R}_\mu$ vanishes in Eq.\eqref{LWZW}, and the WZW term could be simply given as
\begin{eqnarray}
&&\bar{\cal L}_{\rm WZW}=-\frac{N_ce^2}{48\pi^2} \epsilon^{\mu\nu\alpha\beta}A_\mu F_{\alpha\beta} \partial_\nu\theta(x).\label{LWZW1}
\end{eqnarray}
Due to the derivatives $\partial_\nu\theta(x)$, the $2\pi$-periodicity with respect to $\theta(x)$ is automatically guaranteed for $\bar{\cal L}_{\rm WZW}$ as should be. Note that the reduced WZW term is exactly the same as that adopted for the study of $\pi^0$ domain wall in Refs.~\cite{Son:2007ny,Brauner:2016pko} when setting $A_0=\mu_{\rm B}$ and $F_{12}=B$.

In total, the explicit form of the Lagrangian density is
\begin{eqnarray}
{\bar{\cal L}\over f_\pi^2}&=&{1\over2}\partial_\mu\theta(x)\partial^\mu\theta(x)+m_\pi^2\cos \theta(x)\nonumber\\
&&-\frac{\tilde{\rho}_5}{4I_2} \epsilon^{\mu\nu\alpha\beta}A_\mu F_{\alpha\beta} \partial_\nu\theta(x),\label{bLag}
\end{eqnarray}
where we have introduced "chiral density" $\tilde{\rho}_5\equiv \frac{N_ce^2I_2}{12\pi^2f_\pi^2}$ with $I_2\equiv{\bf{E}\cdot{B}}=EB$ the second Lorentz invariant of EM field. The Lagrangian is gauge invariant when $\theta(x)$ varies together with $A_\mu$~\cite{Witten:1983tw}. In order to stabilize the vacuum for the following studies, we take $eE\lesssim m_\pi^2$ to suppress Schwinger pair production and tune $I_2$ simply through $B$. In some work, a constant $\theta$ is introduced to the Lagrangian to study the CP violation of QCD explicitly~\cite{Mameda:2014cxa}. In our consideration, $\theta(x)$ is kind of order parameter that has to be determined by the minimizing the Hamiltonian density
\bea
{{\cal H}\over f_\pi^2}&=&\partial_0\theta{\partial {\bar{\cal L}\over f_\pi^2}\over\partial (\partial_0\theta)}+\partial_0A_\alpha{\partial {\bar{\cal L}\over f_\pi^2}\over\partial (\partial_0A_\alpha)}-{\bar{\cal L}\over f_\pi^2}\nonumber\\
&=&{1\over2}\left[(\partial_0\theta)^2+({\nabla}\theta)^2\right]-m_\pi^2\cos \theta\nonumber\\
&&\ \ \ \ \ \ \ \ \ \ \ \ +\frac{\tilde{\rho}_5}{2I_2} \epsilon^{\mu\nu\alpha\beta}A_\mu \nabla_\alpha A_{\beta} \nabla_\nu\theta(x)\label{Ham}
\eea
with $\nabla_{0}=0$ and $\nabla_i=\partial_i\ (i=1,2,3)$. The minimizing process is equivalent to follow the Euler-Lagrangian equation 
\bea
\partial_\mu{\partial\bar{\cal L}\over\partial \partial_\mu\theta(x)}-{\partial\bar{\cal L}\over\partial \theta(x)}=0,
\eea 
and the static equation of motion (EOM) of $\theta(x)$ can be obtained as
\begin{eqnarray}
	-\boldsymbol{\nabla}^2\theta(x)+m_\pi^2\sin \theta(x)-\tilde{\rho}_5=0.\label{EOM}
\end{eqnarray}
The EOM \eqref{EOM} is invariant with respect to the gauge change of $A_\mu$, we can set $A_\mu$ to be time independent in order to simply reproduce the EOM from Eq.\eqref{Ham}.

If $\theta(x)$ is almost a constant, the Hamiltonian density Eq.\eqref{Ham} becomes
\bea
{{\cal H}\over f_\pi^2}=-m_\pi^2\cos \theta+\frac{\tilde{\rho}_5}{2I_2} \epsilon^{\mu\nu\alpha\beta}A_\mu \nabla_\alpha A_{\beta} \nabla_\nu\theta(x)
\eea
to next-leading order and the EOM follows correspondingly as
\bea
m_\pi^2\sin \theta(x)-\tilde{\rho}_5=0.\label{EOMC}
\eea
For a small $I_2$ or $\tilde{\rho}_5$, two solutions can be obtained from Eq.\eqref{EOMC}, that is,
\bea
\theta_1=\arcsin{\tilde{\rho}_5\over m_\pi^2}, \theta_2=\pi-\theta_1.\label{theta12}
\eea
Surely, $\theta_1$ is a local minimum of ${\cal H}$ and $\theta_2$ a local maximum, so $\theta_1$ is the physical solution and the energy density is
\bea
{{\cal H}\over f_\pi^2}=-m_\pi^2\cos \theta_1.\label{H1}
\eea
We would like to point out that $\theta=2n\pi\ (n\in \mathbb{Z})$ seems to give the lowest energy for any $I_2$, but chiral anomaly drives the system to located at $\theta_1$ instead with extra energy injected from the EM field. However, when $\tilde{\rho}_5>m_\pi^2$, there is no solution to Eq.\eqref{EOMC} at all. This is of course nonphysical and was called "chiral phase instability" in our previous works~\cite{Cao:2015cka,Cao:2019hku,Cao:2020pjq}. Specifically, at the critical point with $\tilde{\rho}_5^c=m_\pi^2$ or $e^2I_2^c={12\pi^2\over N_c}f_\pi^2m_\pi^2$, $\theta_1=\theta_2={\pi\over2}$ and the energy density can be evaluated as ${{\cal H}}^c=0$.

In the following, we will confine ourselves to the super-critical case with $\tilde{\rho}_5>m_\pi^2$ to discuss the "chiral phase instability" problem. When no constant solution can be found to the EOM, it is natural to ask if a space-dependent solution is possible. To answer the question, we have to refer to the general EOM, Eq.\eqref{EOM}.  For demonstration, two shapes of system will be explored: a spherical system in Sec.\ref{SS} and a cylindrical system in Sec.\ref{CS}

\subsection{A spherical system}\label{SS}
\subsubsection{Chiral limit}\label{CL}
We will start with the chiral limit $m_\pi=0$, which is simpler but still contains the most important physical ingredient --  the WZW term. In this case, $I_2^c=0$, and the EOM Eq.\eqref{EOM} is reduced to
\begin{eqnarray}
-\boldsymbol{\nabla}^2\theta(x)=\tilde{\rho}_5.\label{EOM0}
\end{eqnarray}
This is just like the forth Maxwell differential equation with a constant charge density, that is, $\boldsymbol{\nabla\cdot E}(x)=-\boldsymbol{\nabla}^2\phi_0(x)=\rho_e$. So we immediately understand that $\tilde{\rho}_5$ plays a role of  reduced chiral charge density. According to classical electromagnetism, the shape of the system boundary strongly affects the spatial dependence of $\theta(x)$. In this section, we consider a spherical system first, then Gauss' theorem can be applied to find 
\bea
{\bf E}_\theta\equiv-\boldsymbol{\nabla}\theta(x)={{\bf r}\over3}\tilde{\rho}_5
\eea
with ${\bf r}={\bf x}+{\bf y}+{\bf z}$, which then gives
\bea
\theta(r)=-{r^2\over6}\tilde{\rho}_5\label{theta0}
\eea
up to a constant. The constant cannot be constrained due to the exact chiral symmetry of the chiral perturbation theory in chiral limit when EM field is absent. Actually, the EOM Eq.\eqref{EOM0} can be simply reduced to
\begin{eqnarray}
	-{d^2[r\theta(r)]\over r dr^2}=\tilde{\rho}_5
\end{eqnarray}
in the spherical case, and one can check that Eq.\eqref{theta0} is truly the solution. 

In such a case, the corresponding Hamiltonian density is 
\bea
{{\cal H}\over f_\pi^2}&=&{1\over2}(\boldsymbol{\nabla}\theta(x))^2+\frac{\tilde{\rho}_5}{2I_2} \epsilon^{\mu\nu\alpha\beta}A_\mu \nabla_\alpha A_{\beta} \nabla_\nu\theta(x)\nonumber\\
&=&{r^2\over18}\tilde{\rho}_5^2+\frac{r^2}{9}\tilde{\rho}_5^2={r^2\over6}\tilde{\rho}_5^2>0,
\eea
where the isotropic symmetry has been applied to the WZW term in the second step. Compared to ${{\cal H}}^c=0$, we can see that the inhomogeneous solution is less favored than the homogeneous one if exists. If we introduce a small real correction $\delta\theta(r)$ to the solution Eq.\eqref{theta0}, that is,  $\theta(r)=-{r^2\over6}\tilde{\rho}_5+\delta\theta(r)$, then we can derive from Eq.\eqref{Ham}:
\bea
{{\cal H}\over f_\pi^2}-{r^2\over6}\tilde{\rho}_5^2={1\over2}[\boldsymbol{\nabla}\delta\theta(r)]^2>0,
\eea
so the solution is a local minimum.

\subsubsection{Real case}\label{RC}
The situation changes a bit when we consider $m_\pi\neq0$, because there is a local minimum Eq.\eqref{theta12} for $I_2<I_2^c$~\cite{Cao:2015cka} and the Hamiltonian Eq.\eqref{H1} is negative. So the  homogeneous solution is stable until $I_2>I_2^c$, whence we have to solve the following space-dependent differential equation:
\begin{eqnarray}
	-\boldsymbol{\nabla}^2\theta(x)+m_\pi^2\sin\theta(x)=\tilde{\rho}_5.\label{SSx}
\end{eqnarray}
Note that $I_2>I_2^c$ is equivalent to the condition $\tilde{\rho}_5>m_\pi^2$. We are going to solve this equation in the following. 

For a spherical system, the EOM Eq.\eqref{SSx} can be reduced to
\bea
-{d^2[r\theta(r)]\over r dr^2}+m_\pi^2\sin\theta(r)=\tilde{\rho}_5.\label{EOMp}
\eea
We have mentioned that the homogeneous solution is $\theta(r)={\pi\over2}$ at $I_2=I_2^c$ due to the chiral rotation~\cite{Cao:2015cka}. To keep the energy continuous across $I_2^c$, one should choose $\theta(0)={\pi\over2}$ for the inhomogeneous phase. Then, we can check that 
\bea
\theta(r)=-{r^2\over6}(\tilde{\rho}_5-m_\pi^2)+{\pi\over2}
\eea
is the asymptotic solution of Eq.\eqref{EOMp} around the center $r\sim 0$, and the boundary conditions follow as: $\theta(0)={\pi\over2}$ and $\theta'(0)=0$. With that, it is easy to solve the EOM Eq.\eqref{EOMp} numerically and obtain an exact and definite result. 

For $r^2\gg \tilde{\rho}_5^{-1}$, it is convenient to redefine the angle function as $\theta(r)\equiv\hat{\theta}(r)-{r^2\over6}\tilde{\rho}_5$ and the EOM becomes
\bea
-{d^2[r\hat{\theta}(r)]\over r dr^2}+m_\pi^2\sin\left[\hat{\theta}(r)-{r^2\over6}\tilde{\rho}_5\right]=0.\label{hEOM}
\eea
If we assume $|\hat{\theta}(r)|\ll |{r^2\over6}\tilde{\rho}_5|$ in the large $r$ limit, the EOM can be approximately presented as
\bea
-{d^2[r\hat{\theta}(r)]\over r dr^2}-m_\pi^2\sin\left({r^2\over6}\tilde{\rho}_5\right)=0.
\eea
Such a differential equation can be solved analytically to get the general solution as
\bea
\hat{\theta}(r)={3m_\pi^2\over \tilde{\rho}_5}{{\rm C}(\tilde{r})+C_1\over\tilde{r} }+C_2\label{hthetas}
\eea
with $\tilde{r}\equiv{\sqrt{\tilde{\rho}_5\over3\pi}}r$ the reduced radius and $C(z)$ the Fresnel integral function. Since $\lim_{r\rightarrow\infty}\hat{\theta}(r)=C_2$, the prescription $|\hat{\theta}(r)|\ll |{r^2\over6}\tilde{\rho}_5|$ is self-consistently satisfied. Thus, we can conclude that $\theta(r)=-{r^2\over6}\tilde{\rho}_5$ is the asymptotic solution of Eq.\eqref{EOMp} for very large $r$. Moreover, the larger is $\tilde{\rho}_5$, the better is the approximate solution, because $m_\pi^2\sin\theta(r) (\leq m_\pi^2)$ is then only a small correction to $\tilde{\rho}_5$.

For a medium radius, we still need to take into account the correction Eq.\eqref{hthetas} to better reproduce the numerical results. Then, by noticing that $C(\tilde{r})$ oscillates around the center value $1/2$ with $\tilde{r}$ varying, the explicit form could be given as
\bea
\hat{\theta}(r)={3m_\pi^2\over \tilde{\rho}_5}{C(\tilde{r})-C(2)\over\tilde{r} }+\theta(r_0)+2\pi,\label{htheta}
\eea
where $\theta(r_0)$ is the value given by numerical calculations at a large distance $r_0=2\sqrt{3\pi/ \tilde{\rho}_5}$ and $C(2)=0.488$. One can easily check that
$\hat{\theta}(r_0)-2\pi=\theta(r_0)$, that is, the approximate solution exactly matches the numerical one at $r_0$. 

\begin{figure}[!htb]
	\begin{center}
		\includegraphics[width=8cm]{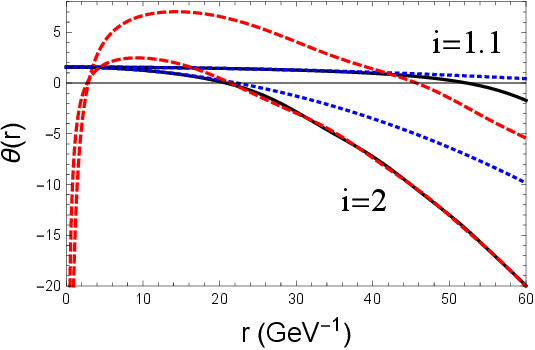}
		\caption{The comparisons between the approximate solutions, $\theta_{\rm s}(r)=-{r^2\over6}(\tilde{\rho}_5-m_\pi^2)+{\pi\over2}$ (blue dotted) and $\theta_{\rm l}(r)=\hat{\theta}(r)-{r^2\over6}\tilde{\rho}_5$ with Eq.\eqref{htheta} (red dashed), and the exact numerical results (black solid) for ${\rm i}\equiv I_2/I_2^c=1.1$ and $2$.}\label{comp}
	\end{center}
\end{figure}

The comparisons between the approximate solutions, $\theta_{\rm s}(r)=-{r^2\over6}(\tilde{\rho}_5-m_\pi^2)+{\pi\over2}$ and $\theta_{\rm l}(r)=\hat{\theta}(r)-{r^2\over6}\tilde{\rho}_5$ with Eq.\eqref{htheta}, and the exact numerical results are given in Fig.\ref{comp} for several $I_2>I_2^c$. As we can see, for $I_2=2 I_2^c$, the numerical results are well reproduced by $\theta_{\rm s}(r)$ and $\theta_{\rm l}(r)$, respectively, in the small and large distance limits. For $I_2\gtrsim I_2^c$,  a large range around $r=0$ can be well reproduced by $\theta_{\rm s}(r)$, but we need a much larger fixed point than $\tilde{r}=2$ for $\theta_{\rm l}(r)$ to better reproduce the large distance limit. So in the limit $I_2\rightarrow I_2^c$ from above, $\theta(r)=-{r^2\over6}(\tilde{\rho}_5-m_\pi^2)+{\pi\over2}$ is the exact solution for a system with a finite radius, and we have
\bea
\lim_{I_2\rightarrow I_2^{c+}}{\theta}(r)&=&{\pi\over2},\\
\lim_{I_2\rightarrow I_2^{c+}} \partial_{I_2}{\theta}(r)&=&-{r^2\over6} \frac{N_ce^2}{12\pi^2f_\pi^2}.
\eea  
Thus, the continuity of energy is guaranteed at $I_2^c$ and the phase transition is of second order.

\begin{figure}[!htb]
	\begin{center}
		\includegraphics[width=8cm]{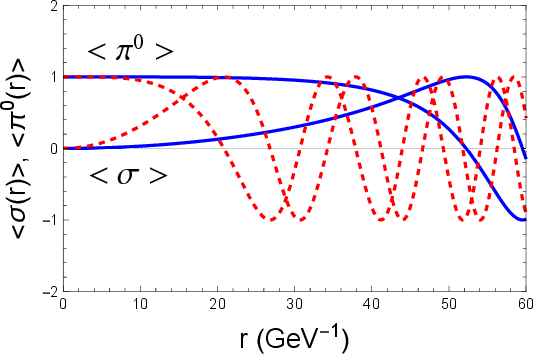}
		\caption{The spatial distributions of reduced condensates $\langle\tilde{\sigma}(r)\rangle\equiv \langle\sigma(r)\rangle/f_\pi$ and $\langle\tilde{\pi}(r)\rangle\equiv \langle\pi(r)\rangle/f_\pi$ for  ${\rm i}\equiv I_2/I_2^c=1.1$ (blue solid) and $2$ (red dashed).}\label{spi}
	\end{center}
\end{figure}

Finally, we show in Fig.\ref{spi} the corresponding evolutions of more physical order parameters,  $\langle\sigma(x)\rangle$ and $\langle\pi^0(x)\rangle$, with the radius $r$. Although $\theta(r)$ is not continuous with $r$ after taking modulus over $2\pi$, $\langle\sigma(x)\rangle$ and $\langle\pi^0(x)\rangle$ are continuous as $2\pi$ are exactly their period with respect to $\theta(r)$. As can be seen, they oscillate with $r$ but the period continuously changes, so such an inhomogeneous phase is very different from the chiral density wave which is periodic in space.

\subsubsection{The nature of phase transition}\label{nature}
What kind of phase transition is it across $I_2^c$? As a matter of fact, neither internal nor spatial symmetry is further broken by the inhomogeneous condensate $\theta(r)$ for the real case, so it is not of Landau type. That can be strongly supported by demonstrating the absence of zero mode fluctuation. If we assume a small fluctuation $\delta\theta(r)$ to the solution $\theta(r)$ in the Lagrangian Eq.\eqref{bLag}, the Lagrangian for the fluctuation field would follow as
\begin{eqnarray}
\delta{{\cal L}}&=&{ f_\pi^2\over2}\partial_\mu\delta\theta(x)\partial^\mu\delta\theta(x)\!-\!f_\pi^2m_\pi^2\cos \theta(x){\delta{\theta}^2(x)\over2}\label{fLag}
\end{eqnarray}
up to order $o(\delta{\theta}^2(x))$. Then, by recalling the spherical symmetry of the system, the EOM of $\delta\theta(x)$ can be given as
\begin{eqnarray}
0&=&\left[\partial_t^2-\partial_r^2-{2\over r}\partial_r\!+\!m_\pi^2\cos \theta(r)\right]{\delta{\theta}(t,r)}\nonumber\\
&=&\left[\partial_t^2-\partial_r^2\!+\!m_\pi^2\cos \theta(r)\right]r{\delta{\theta}(t,r)}.
\end{eqnarray}
For a system with a large radius $R$, we have numerically checked that the zero mode solution would diverge as $r^{-2}$ around $r\sim0$ for nontrivial boundary conditions $\delta{{\theta}(t,R)}=0$ and $\partial_R\delta{{\theta}(t,R)}\neq0$. Such a divergent solution cannot be a candidate for wave function, so there is no massless Nambu-Goldstone mode at all in the inhomogeneous phase. However, since $\cos \theta(r)\sim0$ for $r\sim0$, the collective excitation is almost massless when $r$ is not large, which means that the effective interaction range becomes much larger when $\pi^0$ is exchanged. While, it is easy to find from Eq.\eqref{fLag} that $\delta\theta(r)$ is massless in chiral limit, which follows the Nambu-Goldstone nature of $\pi^0$ and indicates no energy cost to change $\theta(0)$. The physics is quite similar to that of photons: Even when finite electric charges are put into the system and a background electric field is then generated, photons remain massless in this system.

Then, the phase transition at $I_2^c$ can only be of topological type according to our knowledge about the category of phase transition. Surely, the chiral charge $\tilde{\rho}_5$ is of topological nature and increases with increasing $I_2$, but how about the topology of the condensate across the critical point $I_2^c$? We may recall the Gauss' law that the electric field flux through a closed surface, $\Phi=\oint_S {\bf E}\cdot d{\bf S}$ , is solely determined by the number of electric charges inside the surface but does not depend on the distribution, so the field flux serves as a topological invariant. In chiral limit, the background field ${\bf E}_\theta$ becomes coordinate dependent if and only if $I_2$ (thus $\tilde{\rho}_5$) is nonzero. In this sense, the presence of finite chiral charges is consistent with a nontrivial background field flux, defined as $\Phi_\theta\equiv\oint_S {\bf E}_\theta\cdot d{\bf S}$, similar to the case with electric field. 

\begin{figure}[!htb]
	\begin{center}
		\includegraphics[width=8cm]{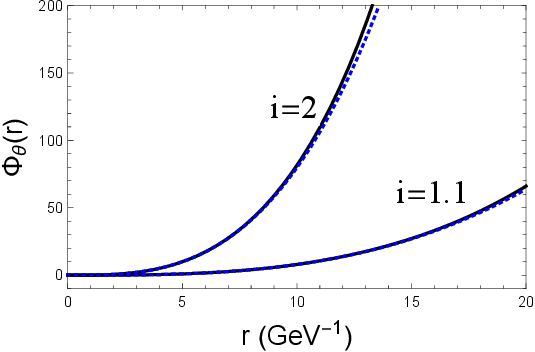}
		\caption{The field flux $\Phi_\theta(r)$ from exact numerical calculations (black solid) and approximate solution $\theta(r)=-{r^2\over6}(\tilde{\rho}_5-m_\pi^2)+{\pi\over2}$ (blue dotted)  for ${\rm i}\equiv I_2/I_2^c=1.1,2$.}\label{Phi}
	\end{center}
\end{figure}

For the real case, the situation is very different: Due to the second term on the right-hand side of Eq.\eqref{bLag} which explicitly breaks chiral symmetry, the flux $\Phi_\theta$ trivially keeps zero up to $I_2^c$ as $\theta(x)$ is homogeneous. It seems that all the chiral charges are completely screened thanks to the reorganization of chiral vacuum. Beyond $I_2^c$, an inhomogeneous profile is developed for $\theta(x)$ and we can evaluate $\Phi_\theta$ for a spherical surface with radius $r$ as
\bea
\Phi_\theta(r)\equiv-4\pi r^2 \partial_r\theta(r).
\eea
Actually, $\Phi_\theta(r)$ depends on the details of chiral charge distribution as the second term is involved in Eq.\eqref{bLag}, so $\Phi_\theta(r)$ can only be understood as an approximate topological order parameter. The exact numerical and approximate results are both presented in Fig.~\ref{Phi} for $\Phi_\theta(r)$. As we can see, the explicit chiral symmetry breaking term seems to contribute a negative charge density $\tilde{\rho}_5'=-m_\pi^2$ for $I_2>I_2^c$. The whole feature is quite similar to that of type-II superconductor in an external magnetic field~\cite{Fetter2003b}: When $B$ is small, it will be completely expelled from the bulk by the superconducting system; but when $B$ becomes larger than the critical one $B_c$, inhomogeneous magnetic vortices would gradually form. So, a chiral charge, such as a deconfined $u/d$ quark, would not feel the field ${\bf E}_\theta$ until $I_2>I_2^c$, when it will get accelerated. However, there is one big difference for the type-II superconductor: the rotational symmetry is spontaneously broken by the magnetic vortex lattice, hence the involved phase transitions are of Landau type.

\subsection{A cylindrical system}\label{CS}
For completeness, we will study a cylindrical system which is infinitely long and infinitely wide in Sec.\ref{IL} and Sec.\ref{IW}, respectively. On one hand, in classical electromagnetism, the electric configuration in such systems can be easily solved when the cylindrical symmetry is consistently taken into account. On the other hand, it is much more convenient to prepare a cylindrical system with constant parallel EM field along its axis in practice. 

\subsubsection{Infinitely long}\label{IL}
We set both the axis of the cylindrical system and the parallel EM field to be along $z$-direction without loss of generality. When the length $L$ along $z$-direction and transverse radius $R$ in $x-y$ plane satisfy $L\gg R$, such a cylindrical system can be approximately considered as infinitely long. It is well-known that the $z$-component of ${\bf E}_\theta$ would vanish in this case, so the EOM Eq.\eqref{EOM} is reduced to
\begin{eqnarray}
\left(\partial_r^2+{1\over r}\partial_r\right)\theta(r)-m_\pi^2\sin \theta(r)+\tilde{\rho}_5=0\label{EOMc}
\end{eqnarray}
with ${\bf r}={\bf x}+{\bf y}$. In chiral limit, it is easy to find
\bea
{\bf E}_\theta={{\bf r}\over2}\tilde{\rho}_5,\ \theta(r)=-{r^2\over4}\tilde{\rho}_5+\theta(0)
\eea
with $\theta(0)$ random. 

By following a similar process as that in Sec.\ref{RC}, if we redefine the field as $\theta(r)=\hat{\theta}(r)-{r^2\over4}\tilde{\rho}_5$, the EOM Eq.\eqref{EOMc} becomes
\begin{eqnarray}
\left(\partial_r^2+{1\over r}\partial_r\right)\hat{\theta}(r)-m_\pi^2\sin \left[\hat{\theta}(r)-{r^2\over4}\tilde{\rho}_5\right]=0.\label{EOMch}
\end{eqnarray}
For $r\sim0$, it can be checked that $\theta(r)=-{r^2\over4}(\tilde{\rho}_5-m_\pi^2)+{\pi\over2}$ is the approximate solution to Eq.\eqref{EOMch}. For a large $r$, the second term is dominated by $-{r^2\over4}\tilde{\rho}_5$ in the sine function, so the EOM can be reduced to
\begin{eqnarray}
\left(\partial_r^2+{1\over r}\partial_r\right)\hat{\theta}(r)+m_\pi^2\sin \left({r^2\over4}\tilde{\rho}_5\right)=0.
\end{eqnarray}
And the general analytic solution is found to be
\begin{eqnarray}
\hat{\theta}(r)=-{m_\pi^2\over \tilde{\rho}_5}\left[{\rm Ci} \left({\tilde{r}^2}\right)+C_1\ln\left({\tilde{r}^2}\right)\right]+C_2
\end{eqnarray}
with ${\rm Ci} (z)$ the cosine integral function and $\tilde{r}={r\over2}\sqrt{\tilde{\rho}_5}$. 

\begin{figure}[!htb]
	\begin{center}
		\includegraphics[width=8cm]{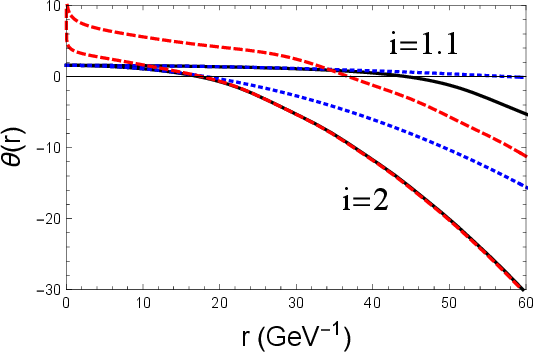}
		\caption{The comparisons between the approximate solutions, $\theta(r)=-{r^2\over4}(\tilde{\rho}_5-m_\pi^2)+{\pi\over2}$ (blue dotted) and $\theta(r)\equiv\hat{\theta}(r)-{r^2\over4}\tilde{\rho}_5$ with Eq.\eqref{hthetac} (red dashed), and the exact numerical results (black solid) for ${\rm i}\equiv I_2/I_2^c=1.1,2$.}\label{compc}
	\end{center}
\end{figure}

Since ${\rm Ci} (z)$ oscillates around $0$ with increasing $z$, the approximate solution can be set to
\begin{eqnarray}
\hat{\theta}(r)&=&-{m_\pi^2\over \tilde{\rho}_5}\left[{\rm Ci} \left({\tilde{r}^2}\right)-{\rm Ci} \left(2.5^2\right)+C_1\ln{\tilde{r}^2\over2.5^2}\right]\nonumber\\
&&+\theta(r_0)+2.5^2,\label{hthetac}
\end{eqnarray}
where $\theta(r_0)$ is the value given by numerical calculations at a large distance $r_0=5/\sqrt{\tilde{\rho}_5}$ and ${\rm Ci} \left(2.5^2\right)=-0.028$. Then, one can check that
$\hat{\theta}(r_0)-2.5^2=\theta(r_0)$, that is, the approximate solution exactly matches the numerical one at $r_0$. The constant $C_1$ is very important for the accuracy of the approximate solution in medium range, since $\ln\left({\tilde{r}^2}\right)$ increases with $r$. The numerical fitting shows that $C_1=-1/2$ can reproduce the exact results very well for larger $I_2$, see Fig.~\ref{compc}. The features are quite similar to those of a spherical system and the transition is still of second order at $I_2^c$.
\subsubsection{Infinitely wide}\label{IW}
To the contrary, if $L\ll R$,  such a cylindrical system can be approximately considered as infinitely wide. It is well-known that only the $z$-component of ${\bf E}_\theta$ is nonzero in this case, so the  EOM Eq.\eqref{EOM} is reduced to
\begin{eqnarray}
\partial_z^2\theta(z)-m_\pi^2\sin \theta(z)+\tilde{\rho}_5=0.\label{EOMc2}
\end{eqnarray}
For $\tilde{\rho}_5=0$, chiral soliton lattice is a solution of this EOM~\cite{Brauner:2016pko}, but it is not the ground state compared to the homogeneous phase with $\theta(z)=0$. In chiral limit, it is easy to find
\bea
{\bf E}_\theta={{\bf z}}\tilde{\rho}_5,\ \theta(z)=-{z^2\over2}\tilde{\rho}_5
\eea
with $\theta(0)$ random. 

By following a similar process as that in Sec.\ref{RC}, if we redefine the field as $\theta(z)=\hat{\theta}(z)-{z^2\over2}\tilde{\rho}_5$, the EOM Eq.\eqref{EOMc2} becomes
\begin{eqnarray}
\partial_z^2\hat{\theta}(z)-m_\pi^2\sin \left[\hat{\theta}(z)-{z^2\over2}\tilde{\rho}_5\right]=0.\label{EOMch2}
\end{eqnarray}
For $z\sim0$, it can be checked that $\theta(r)=-{z^2\over2}(\tilde{\rho}_5-m_\pi^2)+{\pi\over2}$ is the approximate solution to Eq.\eqref{EOMch2}. For a large $|z|$, the second term is dominated by $-{z^2\over2}\tilde{\rho}_5$ in the sine function, so the EOM can be reduced to
\begin{eqnarray}
\partial_z^2\hat{\theta}(z)+m_\pi^2\sin \left({z^2\over2}\tilde{\rho}_5\right)=0.
\end{eqnarray}
And the general analytic solution is found to be
\begin{eqnarray}
\hat{\theta}(z)={m_\pi^2\over \tilde{\rho}_5}\left[\cos \left({{\pi\over2}\tilde{z}^2}\right)+\pi \tilde{z}\,{\rm S}(\tilde{z})-C_1 \tilde{z}\right]+C_2
\end{eqnarray}
with ${\rm S} (z)$ the Fresnel integral function and $\tilde{z}=|z|\sqrt{\tilde{\rho}_5/\pi}$. 

\begin{figure}[!htb]
	\begin{center}
		\includegraphics[width=8cm]{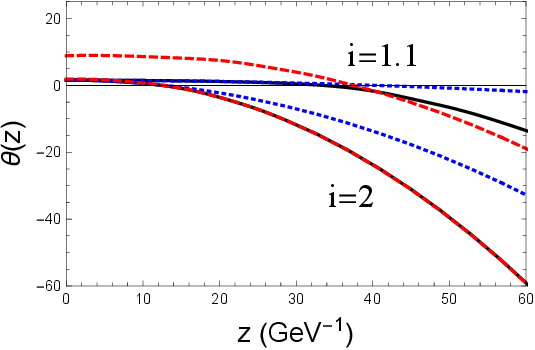}
		\caption{The comparisons between the approximate solutions,  $\theta(z)=-{z^2\over2}(\tilde{\rho}_5-m_\pi^2)+{\pi\over2}$ (blue dotted) and $\theta(z)\equiv\hat{\theta}(z)-{z^2\over2}\tilde{\rho}_5$ with Eq.\eqref{hthetac2} (red dashed), and the exact numerical results (black solid) for ${\rm i}\equiv I_2/I_2^c=1.1,2$.}\label{compc2}
	\end{center}
\end{figure}

Since $\cos \left({{\pi\over2}\tilde{z}^2}\right)$ and ${\rm S}(\tilde{z})$ oscillate around $0$ and $1/2$, respectively, we would $C_2$ at $\tilde{z}^2_0=11$ where $\cos \left({{\pi\over2}\tilde{z}^2_0}\right)=0$ and ${\rm S}(\tilde{z}_0)=0.503$. We have
\begin{eqnarray}
\hat{\theta}(z)&=&{m_\pi^2\over \tilde{\rho}_5}\Big\{\cos \left({{\pi\over2}\tilde{z}^2}\right)+\pi \tilde{z}\,\left[{\rm S}(\tilde{z})-{\rm S}(\tilde{z}_0)\right]\nonumber\\
&&-C_1\left(\tilde{z}-\tilde{z}_0\right)\Big\}+\theta(z_0)+{11\over2}\pi,\label{hthetac2}
\end{eqnarray}
where $\theta(z_0)$ is the value given by numerical calculations at a large distance $z_0=\sqrt{11\pi/\tilde{\rho}_5}$. Then, one can check that
$\hat{\theta}(z_0)-{11\over2}\pi=\theta(z_0)$, that is, the approximate solution exactly matches the numerical one at $z_0$. 
The constant $C_1$ is very important for the accuracy of the approximate solution in medium range, since $\tilde{z}$ increases with $z$. The numerical fitting shows that $C_1=-\sqrt{6}$ can reproduce the exact results very well over the whole range for larger $I_2$, see Fig.~\ref{compc2}. Again, the features are quite similar to those of a spherical system and the transition is still of second order at $I_2^c$.

\section{Three-flavor chiral perturbation theory }\label{3f}
The Lagrangian of two-flavor chiral perturbation theory can be easily extended to the three-flavor case by redefining the unitary matrix $U$ in $SU(3)$ flavor space, that is,
\bea
U=\exp\left(i \sum_{a=1}^8\lambda^a\phi^a(x)\right)
\eea
with $ \lambda^a$ Gell-Mann matrices and $\phi^a$ the corresponding octuplet pseudoscalar fields. The Lagrangian is now
\begin{eqnarray}
\label{chirall}
{\cal L}={\cal L}_0+{\cal L}_{\rm WZW},
\end{eqnarray}
where the normal chiral Lagrangian ${\cal L}_0$ changes to~\cite{Scherer:2002tk}
\begin{eqnarray}
{\cal L}_0=\frac{f_\pi^2}{4}\tr\left[ D_\mu U^\dagger D^\mu U+{m_\pi^2}(\tilde{M}^\dagger U+U^\dagger \tilde{M})\right]\label{L30}
\end{eqnarray}
and the Wess-Zumino-Witten term ${\cal L}_{\rm WZW}$ takes the same form as Eq.\eqref{LWZW}.
In Eq.\eqref{L30}, $\tilde{M}={\rm diag}(1,1,r_s)$ is the reduced mass matrix with $r_s\equiv m_s/m=24.3$ the ratio between $s$ and $u/d$ quark masses~\cite{Scherer:2002tk}; the covariant derivative is defined as
\begin{eqnarray}
D_\mu U=\partial_\mu U+A_\mu[Q,U],
\end{eqnarray}
where $Q={\rm diag}(2/3,-1/3,-1/3)e$ is the charge matrix in $SU(3)$ flavor space. In the limit $m_s\rightarrow\infty$, only the lightest pseudoscalars $\boldsymbol{\pi}\propto(\phi^1,\phi^2,\phi^3)$ are effective, and the theory exactly reduces to the two-flavor one with the Lagrangian given by Eq.\eqref{L0}.

For the three-flavor case, since $\phi^8$ is neutral and takes part in the triangle anomaly, it might condense as $\pi^0$ does in the presence of a parallel EM field. If we set only $\phi^3$ and $\phi^8$ to be nonzero in $U$, the unitary matrix can be simply reduced to
\bea
U={\rm diag}\left(e^{i\,\left(\phi^3+{\phi^8\over\sqrt{3}}\right)},e^{i\,\left(-\phi^3+{\phi^8\over\sqrt{3}}\right)},e^{-i\,{2\phi^8\over\sqrt{3}}}\right).\label{U3}
\eea
From such a form, we can identify 
\bea
\langle\sigma_u\rangle&=&f_\pi\cos\left(\phi^3+{\phi^8\over\sqrt{3}}\right), 
\langle\sigma_d\rangle=f_\pi\cos\left(-\phi^3+{\phi^8\over\sqrt{3}}\right), \nonumber\\
\langle\sigma_s\rangle&=&f_\pi\cos{2\phi^8\over\sqrt{3}},
\langle\pi_u\rangle=f_\pi\sin\left(\phi^3+{\phi^8\over\sqrt{3}}\right), \nonumber\\
\langle\pi_d\rangle&=&f_\pi\sin\left(-\phi^3+{\phi^8\over\sqrt{3}}\right), 
\langle\pi_s\rangle=f_\pi\sin{2\phi^8\over\sqrt{3}},\label{ops}
\eea
which are consistent with Eq.\eqref{sp} in the limit $\phi^8\rightarrow0$. But for $\phi^8\neq0$, the eigenstates are not $\pi^0$ and $\eta$ any more due to flavor mixing~\cite{Cao:2020pjq} . By substituting Eq.\eqref{U3} into Eq.\eqref{L30}, we find the explicit form of the normal chiral Lagrangian as
\begin{eqnarray}
{\bar{\cal L}_0\over f_\pi^2}\!&=&\!{1\over2}\!\!\sum_{a=3,8}\!\!\partial_\mu\phi^a\partial^\mu\phi^a\!+\!m_\pi^2\left[\cos \phi^3\cos {\phi^8\over\sqrt{3}}\!+\!r_s\cos^2{\phi^8\over\sqrt{3}}\right]\nonumber\\
\end{eqnarray}
up to an irrelevant constant. When $\phi^3,\phi^8\sim0$, we can check that their masses are $m_\pi$ and $m_\eta\equiv\sqrt{1+2r_s\over3}m_\pi$, respectively. So, we can fix $r_s=24.3$ by applying the physical masses of $\pi^0$ and $\eta$ mesons~\cite{Scherer:2002tk}.

To evaluate the WZW term Eq.\eqref{LWZW} in the three-flavor case, we note that
\bea
\bar{L}_\mu=\bar{R}_\mu=-i\,\sum_{a=3,8}\lambda^a\,\partial_\mu\phi^a(x),
\eea
which then implies the EV of $T_\nu$ to be
\bea
\bar{T}_\nu=3\,\tr\, Q^2\bar{L}_\nu=-i\,e^2\partial_\nu\left[\phi^3(x)+{\phi^8(x)\over\sqrt{3}}\right].
\eea
So, the total Lagrangian can be given explicitly as
\begin{eqnarray}
{\bar{\cal L}\over f_\pi^2}&=&{\bar{\cal L}_0\over f_\pi^2}-\frac{\tilde{\rho}_5}{4I_2} \epsilon^{\mu\nu\alpha\beta}A_\mu F_{\alpha\beta} \partial_\nu\left[\phi^3(x)+{\phi^8(x)\over\sqrt{3}}\right],\label{bLag3}
\end{eqnarray}
where we find that the chiral charge density involved for $\phi^8(x)$ is ${1\over\sqrt{3}}$ of that for $\phi^3(x)$ for a given $I_2$. Recalling that $m_{\eta}\gg m_\pi$, we expect the two-flavor results presented in Sec.\ref{2f} to be very conceivable when $m_\pi^2\ll\tilde{\rho}_5\ll m_\eta^2$.

To study the ground state, we apply the Euler-Lagrangian equation 
\bea
\partial_\mu{\partial\bar{\cal L}\over\partial \partial_\mu\phi^a(x)}-{\partial\bar{\cal L}\over\partial \phi^a(x)}=0
\eea 
and derive the static EOMs of $\phi^3(x)$ and $\phi^8(x)$ as
\begin{eqnarray}
\left\{\begin{array}{l}
	-\boldsymbol{\nabla}^2\phi^3+m_\pi^2\sin \phi^3\cos {\phi^8\over\sqrt{3}}=\tilde{\rho}_5,\\
	-\boldsymbol{\nabla}^2\phi^8+{m_\pi^2\over\sqrt{3}}(\cos \phi^3+2r_s\cos {\phi^8\over\sqrt{3}})\sin {\phi^8\over\sqrt{3}}={\tilde{\rho}_5\over\sqrt{3}},
\end{array}\right.\label{EOMs}
\end{eqnarray}
which are coupled differential equations. For $I_2\leq I_2^c$ or $\tilde{\rho}_5\leq m_\pi^2$, homogeneous phase is expected and the EOMs Eq.\eqref{EOMs} reduce to
\begin{eqnarray}
\left\{\begin{array}{l}
	m_\pi^2\sin \phi^3\cos {\phi^8\over\sqrt{3}}=\tilde{\rho}_5,\\
	{m_\pi^2\over\sqrt{3}}(\cos \phi^3+2r_s\cos {\phi^8\over\sqrt{3}})\sin {\phi^8\over\sqrt{3}}={\tilde{\rho}_5\over\sqrt{3}}.
\end{array}\right.\label{EOMsh}
\end{eqnarray}
As $\tilde{\rho}_5\ll m_\eta^2$ in this regime, $\phi^8$ is small and these algebra equations can be approximately solved to give
\bea
\sin\phi^3={\tilde{\rho}_5\over m_\pi^2}, \ \sin{\phi^8\over\sqrt{3}}={\sin\phi^3\over \cos \phi^3+2r_s}.
\eea
We can check that $\sin{\phi^8\over\sqrt{3}}\approx {1\over 48.6}\ll 1$ at $I_2^c$, so the approximation is truly very good. For $\tilde{\rho}_5>m_\pi^2$, the first equation of Eq.\eqref{EOMsh} cannot be satisfied at all, so $I_2^c$ remains a critical point in the three-flavor case. Similarly, there could be another critical point $I_2^{c'}$ induced by the presence of $\eta$ meson, beyond which the second equation of Eq.\eqref{EOMsh} can never be satisfied for any given $\phi^3$. Actually, $\phi^3$ is not important for the determination of $I_2^{c'}$ as $r_s\gg1$, and the critical point can be easily found to be
\bea
I_2^{c'}\approx r_s I_2^c
\eea
when ${\tilde{\rho}_5\over\sqrt{3}}$ reaches the maximum of the left-hand side.

 \subsection{Inhomogeneous phase}
 In the following, we will take a cylindrical system of infinitely long for example to show how inhomogeneous phase develops in the case $I_2>I_2^c$. Similar to Eq.\eqref{EOMc} in Sec.\ref{IL}, the EOMs can be reduced to
 \begin{eqnarray}
\left\{\!\!\begin{array}{l}
	\left(\partial_r^2+{1\over r}\partial_r\right)\phi^3-m_\pi^2\sin \phi^3\cos {\phi^8\over\sqrt{3}}+\tilde{\rho}_5=0,\\
	\left(\partial_r^2\!+\!{1\over r}\partial_r\right)\phi^8\!-\!{m_\pi^2\over\sqrt{3}}(\cos \phi^3\!+\!2r_s\cos {\phi^8\over\sqrt{3}})\sin {\phi^8\over\sqrt{3}}\!+\!{\tilde{\rho}_5\over\sqrt{3}}\!=\!0
\end{array}\right.\label{EOMsc}
\end{eqnarray}
with ${\bf r}={\bf x}+{\bf y}$. Since now $\phi^3$ develops an inhomogeneous feature from the first equation of Eq.\eqref{EOMsc}, it follows from the second equation that $\phi^8$ must also be inhomogeneous. In chiral limit $m_\pi\rightarrow0$, the EOMs decouple in Eq.\eqref{EOMsc}, and we simply have the solutions:
 \bea
 \phi^3=-{r^2\over4}\tilde{\rho}_5+C_1, \ \phi^8=-{r^2\over4\sqrt{3}}\tilde{\rho}_5+C_2.
 \eea
For the physical case, the approximate solution of $\phi^3$ follows as
 \bea
 \phi^3&=&-{r^2\over4}\left(\tilde{\rho}_5-m_\pi^2\cos {\phi^8\over\sqrt{3}}\right)+{\pi\over2}\label{phi30}
  \eea
 around $r\sim0$ up to $o(r^2)$. But for $I_2<I_2^{c'}$, that of $ \phi^8$ takes the constant solution
 \bea
 \phi^8&=&{\sqrt{3}\over2}\arcsin{\tilde{\rho}_5\over r_sm_\pi^2}\label{phi80}
 \eea
by substituting Eq.\eqref{phi30} into the second equation of Eq.\eqref{EOMsc}. One can easily check that Eqs.\eqref{phi30} and \eqref{phi80} would consistently reproduce the homogeneous phase at $\tilde{\rho}_5=m_\pi^2$, hence the energy is continuous across $I_2^c$ and the transition remains of second order. Eventually, the initial conditions follow as
 \bea
&& \phi^3(0)={\pi\over2},  {\phi^3}'(0)=0,\label{IC1}\\
&& \phi^8(0)={\sqrt{3}\over2}\arcsin{\tilde{\rho}_5\over r_sm_\pi^2},\ {\phi^8}'(0)=0\label{IC2}
 \eea
 for $I_2^c<I_2<I_2^{c'}$, but
  \bea
 \phi^3(0)={\pi\over2},  {\phi^3}'(0)=0, \phi^8(0)={\sqrt{3}\pi\over4},\ {\phi^8}'(0)=0\label{IC3}
 \eea
 for $I_2>I_2^{c'}$. Again, the continuity of initial conditions across $I_2^{c'}$ guarantees the continuity of energy, and the transition can be shown to be of second order by following a similar discussion as in Sec.\ref{RC}.
  
\begin{figure}[!htb]
	\begin{center}
		\includegraphics[width=8cm]{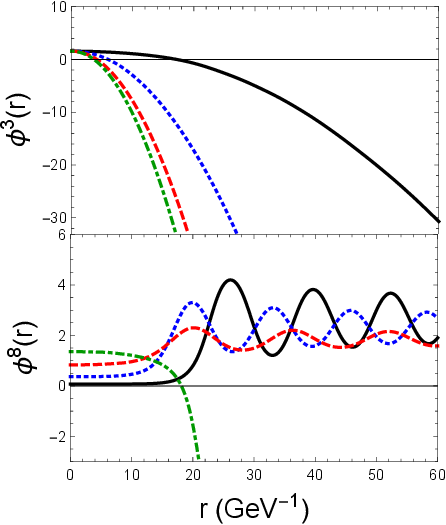}
		\caption{The coordinate dependences of $\phi^3(r)$ (upper panel) and $\phi^8(r)$ (lower panel) for $I_2/I_2^c=2$ (black solid lines), $10$ (blue dotted lines), $20$ (red dashed lines) and $25$ (green dashdotted lines).}\label{phi38}
	\end{center}
\end{figure}

With the initial conditions Eqs.\eqref{IC1}, \eqref{IC2} and \eqref{IC3}, the EOMs Eq.\eqref{EOMsc} can be solved definitely, and the numerical results are illustrated in Fig.\ref{phi38} for several $I_2$. In the upper panel, the result of $\phi^3(r)$ is quite consistent with that given in Fig.\ref{compc} for $I_2=2I_2^c$. Actually, the larger is $I_2$, the better is the consistency, since then the mass term is less important. However, for the case $I_2\sim I_2^c$ (not shown), the first equation of Eq.\eqref{EOMsc} is very sensitive to the presence of $\phi^8(r)$, and $\phi^3(r)$ decreases more quickly than that given in Fig.\ref{compc}. In the lower panel, it is interesting to notice that $\phi^8(r)$ develops an oscillating feature for $I_2<I_2^{c'}$ but decreases very quickly for 
$I_2>I_2^{c'}$. Recalling the discussions in Sec.\eqref{nature}, the oscillating feature actually implies a lattice structure of the effective field $E_8\equiv-\partial_r\phi^8(r)$, so the transition is also topological at $I_2^{c'}$. We will show in more detail later that the oscillation center of ${\phi^8(r)\over\sqrt{3}}$ is at $\sim {\pi\over2}$ for $I_2\sim I_2^c$; so the screening chiral charge, $m_\pi^2\cos {\phi^8\over\sqrt{3}}$, vanishes on average for larger $r$. As a consequence, the effective chiral charge for $\phi^3(r)$ is $\tilde{\rho}_5$ at larger $r$ instead of $\tilde{\rho}_5-m_\pi^2\sim0$ at $r\sim0$ -- that explains why $\phi^3(r)$ decreases more quickly for $I_2\sim I_2^c$ in the three-flavor case.  

Now, we pay attentions to the oscillating feature of $\phi^8(r)$ at larger $r$. Without the mixing term $\propto\cos \phi^3\sin {\phi^8\over\sqrt{3}}$, it is obvious that $\phi^8(r)=\phi^8(0)$ is the exact solution of the second equation of Eq.\eqref{EOMsc}, so the spatial dependence of $\phi^3(r)$ is the key for developing the oscillation. We have mentioned that the mass term for $\phi^3$ is not important when $\tilde{\rho}_5\gg m_\pi^2$, so the solution of the first equation of Eq.\eqref{EOMsc} can be simply given by Eq.\eqref{phi30} that automatically satisfies the constraints at $r\sim0$. Then, the second equation of Eq.\eqref{EOMsc} can be reduces to
\bea
&&\!\left(\!\partial_r^2\!+\!{1\over r}\partial_r\!\right)\phi^8\!-\!{m_\pi^2\over\sqrt{3}}\!\left[\sin {\phi^3}'\!+\!2r_s\cos {\phi^8\over\sqrt{3}}\right]\!\sin {\phi^8\over\sqrt{3}}\!+\!{\tilde{\rho}_5\over\sqrt{3}}\!=\!0\nonumber\\
\label{EOM80}
\eea
with ${\phi^3}'(r)\equiv {r^2\over4}(\tilde{\rho}_5-m_\pi^2)$. In Sec.\ref{2f}, we have shown that there are two independent homogeneous solutions, Eq.\eqref{theta12}, to the EOM of $\theta$ for $I_2<I_2^c$. Similarly, the two independent homogeneous solutions for $\phi^8$ are
\bea
\phi^8_0\equiv\phi^8(0),{\sqrt{3}\over 2}\pi-\phi^8_0
\eea
 for $I_2<I_2^{c'}$ if $\phi_3={\pi/ 2}$. The solution $\phi^8_0$ is a local minimum and serves as a self-consistent initial condition, while ${\sqrt{3}\over 2}\pi-\phi^8_0$ is a local maximum. Actually, we have checked that the latter is exactly the oscillation center for larger $r$ and gradually approaches $\phi^8_0$ as $I_2\rightarrow I_2^{c'}$.

If we redefine $\phi^8(r)\equiv\hat{\phi}^8(r)+{\sqrt{3}\over 2}\pi-\phi^8(0)$, Eq.\eqref{EOM80} can be reduced to
\bea
&&\left(\partial_r^2\!+\!{1\over r}\partial_r\right)\hat{\phi}^8\!-\!{m_\pi^2\over3}\left\{\sin {\phi^3}'\sin{\phi^8_0\over\sqrt{3}}-2r_s\cos {2\phi^8_0\over\sqrt{3}}\right\}\hat{\phi}^8\nonumber\\
&&-{m_\pi^2\over\sqrt{3}}\sin {\phi^3}'\cos{\phi^8_0\over\sqrt{3}}\!=\!0. \label{EOM81}
\eea
 up to $o(\hat{\phi}^8)$. Here, we find that the chiral charge ${\tilde{\rho}_5\over\sqrt{3}}$ is completely cancelled out by the reorganization of vacuum around $\phi^8={\sqrt{3}\over 2}\pi-\phi^8_0$. Instead, there is an effective chiral charge oscillating over space, see the last term on the left-hand side of Eq.\eqref{EOM81}. So the existence of a homogeneous solution to the second equation of Eq.\eqref{EOMsc} for ${\phi^3}={\pi\over2}$ is important for developing a lattice structure -- this explains the topological change of $\phi^8(r)$ across $I_2^{c'}$. In fact, the effect of the ${\phi^3}'$ dependent terms is to catalyze the evolution of  $\phi^8(r)$ from $\phi^8_0$ to around ${\sqrt{3}\over 2}\pi-\phi^8_0$; after that, they can be roughly neglected for larger $r$ as $r_s\gg1$. Then, Eq.\eqref{EOM81} can be simply reduced to
 \bea
\left(\partial_r^2\!+\!{1\over r}\partial_r\right)\hat{\phi}^8\!+\!\tilde{m}_\eta^2\hat{\phi}^8=0, \label{EOM82}
\eea
where the effective mass of $\eta$ meson is
\bea
\tilde{m}_\eta^2\equiv{2r_s\over3}\cos {2\phi^8_0\over\sqrt{3}}m_\pi^2={2\over3}\sqrt{(r_sm_\pi^2)^2-\tilde{\rho}_5^2}
\eea
after substituting the explicit value of $\phi^8_0$. Note that $\tilde{m}_\eta$ vanishes at $I_2^{c'}$ indicating a phase transition.

\begin{figure}[!htb]
	\begin{center}
		\includegraphics[width=8cm]{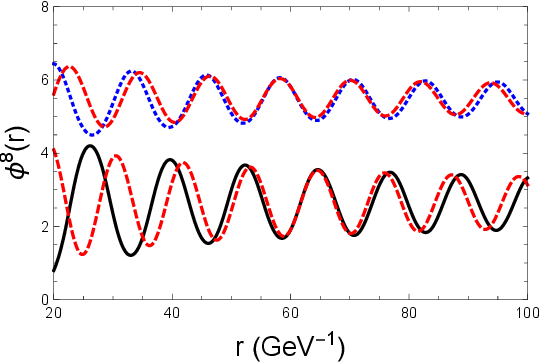}
		\caption{The comparisons between the approximate solution Eq.\eqref{phi8a} (red dashed lines) and exact numerical results (black solid and blue dotted lines) of $\phi^8(r)$ for $I_2/I_2^c=2$ and $10$. For clarity, the results for $I_2/I_2^c=10$ are shifted up by $\pi$. The constants $C_1$ and $C_2$ are fixed by fitting to two data points at $r=60$ and $70\ {\rm GeV}^{-1}$.}\label{comp8}
	\end{center}
\end{figure}

The general solution of Eq.\eqref{EOM82} is
\bea
{\hat{\phi}^8\over\sqrt{3}}=C_1J_0(\tilde{m}_\eta r)+C_2Y_0(\tilde{m}_\eta r),
 \eea
 where $J_0(z)$ and $Y_0(z)$ are Bessel functions of the first and second kinds, respectively. So, the wave length of the oscillation is found to be $2\pi/\tilde{m}_\eta$, and the approximate solution of ${\phi}^8$ is
\bea
\!\!\!\!\!{\phi^8\over\sqrt{3}}\!=\!C_1J_0(\tilde{m}_\eta r)\!+\!C_2Y_0(\tilde{m}_\eta r)\!+\!{\pi\!-\!\arcsin{\tilde{\rho}_5\over r_sm_\pi^2}\over 2},\label{phi8a}
\eea
where the constants $C_1$ and $C_2$ can be fixed by fitting to two data points at larger $r$. The comparisons with exact numerical results are presented in Fig.~\ref{comp8} for different $I_2$. As we can see, the main features are captured by the approximate solution Eq.\eqref{phi8a} at larger $r$: the oscillation center is located at ${\sqrt{3}\over 2}\pi-\phi^8_0$, the wave-length is approximately $2\pi/\tilde{m}_\eta$, and the damping of the amplitude follows that of Bessel functions. Furthermore, the approximate solution works better when $I_2$ is larger.

\begin{figure}[!htb]
	\begin{center}
		\includegraphics[width=8cm]{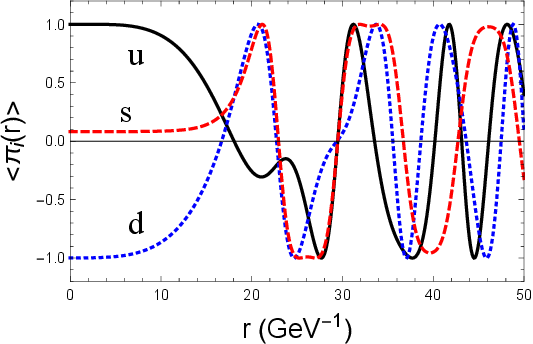}
		\caption{The evolutions of the pseudoscalar order parameters $ \langle{\pi}_i\rangle\ (i=u,d,s)$ (reduced by $f_\pi$) for $I_2/I_2^c=2$. The black solid, blue dotted and red dashed  lines correspond to $u,d$ and $s$ components, respectively.}\label{pops}
	\end{center}
\end{figure}

Finally, take the case $I_2=2I_2^c$ for example, we present the more physical pseudoscalar order parameters in Fig.\ref{pops} by following the definitions in Eq.\eqref{ops}. As can be seen,
\bea
|\langle\pi_{u/d}\rangle|\sim f_\pi, \langle\pi_s\rangle\sim 0
\eea
at the origin, so chiral rotation has been completed for the $u/d$ quark sector but is still very small for the $s$ quark sector. Nevertheless, they all oscillate at larger $r$ when inhomogeneous effect becomes significant. In fact, the wave-length of $\langle\pi_{u/d}\rangle$ decreases very quickly with increasing $I_2\ (<I_2^{c'})$, but that of $\langle\pi_s\rangle$ does not change much by following the lattice structure of $\phi^8$.

\section{Summary and prospects}\label{sum}
In this work, we attack the problem of “chiral phase instability” under a parallel and constant electromagnetic field within the two- and three-flavor chiral perturbation theory. In Sec.\ref{2f}, an effective Lagrangian density is obtained for the chiral angle within the two-flavor ChPT and the WZW term is found to play a role of chiral source. In the case $I_2>I_2^c$, we discover that the $\chi$PI actually implies an inhomogeneous phase, where $\theta(r)$ is proportional to $r^2$ when the relevant spatial scale $r$ or the chiral charge $\tilde{\rho}_5$ is large. The physics is similar to that of an electric field produced in a system with homogeneous distribution of electric charge density. But one should be very careful with the results at large $r$ as $\partial_r\theta$ might exceed the valid energy scale of ChPT, $4\pi f_\pi$, then higher-order corrections are needed. According to our detailed discussions in Sec.\ref{nature}, the phase transition is of topological and second order at $I_2^c$ with $\Phi_\theta$ the topological order parameter. To the best of our knowledge, this is the first time a pure topological transition is discovered in a realistic QCD system. In Sec.\ref{3f}, the work is extended to the three-flavor case, where octuplet pseudoscalars are involved instead of the triplet pions in the two-flavor case. As expected, an inhomogeneous $\eta$ condensation is also found to be developed together with $\pi^0$ for $I_2>I_2^c$. Correspondingly, there is a second critical point, $I_2^{c'}=r_sI_2^c \ (r_s=24.3)$, across which the transition is also of topological and second order by its very nature.

However, this is not the end of story at least for two reasons: 1. The chiral symmetry breaking and restoration is not self-consistently taken into account along with chiral rotation. 2. The effect of singlet pseudoscalar $\eta'$ has not been considered. According to our previous studies in Nambu--Jona-Lasinio model, the effect of chiral symmetry restoration is small around $I_2^c$ but would become significant with a larger $I_2$~\cite{Cao:2015cka}. In this sense, the second critical point $I_2^{c'}$ is not so reliable. Furthermore, a large $I_2$ would induce remarkable flavor mixing among $\pi^0, \eta$ and $\eta'$ mesons~\cite{Cao:2020pjq}, so the EOMs will be more involved and the features of pseudoscalar condensations become more complicated. To understand these effects more physically, we will extend the study to both two- and three-flavor Nambu--Jona-Lasinio models in the future.

\section*{Acknowledgement}
G. Cao thanks Xu-guang Huang for helpful discussions. G. C. is supported by the Natural Science Foundation of Guangdong Province.


\begin{thebibliography}{20}
\bibitem{Kharzeev:2022hqz}
D.~E.~Kharzeev, J.~Liao and S.~Shi,
``Implications of the isobar-run results for the chiral magnetic effect in heavy-ion collisions,''
Phys. Rev. C \textbf{106}, L051903 (2022).

\bibitem{Chen:2023jhx}
B.~X.~Chen, X.~L.~Zhao and G.~L.~Ma,
``On the difference between signal and background of the chiral magnetic effect relative to spectator and participant planes in isobar collisions at $\sqrt{s_{_{\rm NN}}} = 200$ GeV,''
[arXiv:2301.12076 [nucl-th]].

\bibitem{Xu:2023elq}
Z.~Xu, B.~Chan, G.~Wang, A.~Tang and H.~Z.~Huang,
``Event Shape Selection Method in Search of the Chiral Magnetic Effect in Heavy-ion Collisions,''
[arXiv:2307.14997 [nucl-th]].

\bibitem{Wang:2023yis}
J.~f.~Wang, H.~j.~Xu and F.~Wang,
``Impact of initial fluctuations and nuclear deformations in isobar collisions,''
[arXiv:2305.17114 [nucl-th]].

\bibitem{STAR:2023gzg}
 [STAR],
``Upper Limit on the Chiral Magnetic Effect in Isobar Collisions at the Relativistic Heavy-Ion Collider,''
[arXiv:2308.16846 [nucl-ex]].

\bibitem{STAR:2021mii}
M.~Abdallah \textit{et al.} [STAR],
``Search for the chiral magnetic effect with isobar collisions at $\sqrt {s_{NN}}$=200 GeV by the STAR Collaboration at the BNL Relativistic Heavy Ion Collider,''
Phys. Rev. C \textbf{105}, 014901 (2022).

\bibitem{Adler:1969gk} 
S.~L.~Adler,
``Axial vector vertex in spinor electrodynamics,''
Phys.\ Rev.\  {\bf 177}, 2426 (1969).

\bibitem{Bell:1969ts} 
J.~S.~Bell and R.~Jackiw,
``A PCAC puzzle: pi0 $\rightarrow$ gamma gamma in the sigma model,''
Nuovo Cim.\ A {\bf 60}, 47 (1969).

\bibitem{Kharzeev:2007jp} 
D.~E.~Kharzeev, L.~D.~McLerran and H.~J.~Warringa,
``The Effects of topological charge change in heavy ion collisions: 'Event by event P and CP violation',''
Nucl.\ Phys.\ A {\bf 803}, 227 (2008).


\bibitem{Fukushima:2008xe} 
K.~Fukushima, D.~E.~Kharzeev and H.~J.~Warringa,
``The Chiral Magnetic Effect,''
Phys.\ Rev.\ D {\bf 78}, 074033 (2008).

\bibitem{Kharzeev:2010gd}
D.~E.~Kharzeev and H.~U.~Yee,
``Chiral Magnetic Wave,''
Phys. Rev. D \textbf{83}, 085007 (2011).

\bibitem{Son:2004tq} 
D.~T.~Son and A.~R.~Zhitnitsky,
``Quantum anomalies in dense matter,''
Phys.\ Rev.\ D {\bf 70}, 074018 (2004).

\bibitem{Metlitski:2005pr} 
M.~A.~Metlitski and A.~R.~Zhitnitsky,
``Anomalous axion interactions and topological currents in dense matter,''
Phys.\ Rev.\ D {\bf 72}, 045011 (2005).

\bibitem{Huang:2013iia} 
X.~G.~Huang and J.~Liao,
``Axial Current Generation from Electric Field: Chiral Electric Separation Effect,''
Phys.\ Rev.\ Lett.\  {\bf 110}, 232302 (2013).

\bibitem{Hattori:2016njk}
K.~Hattori and Y.~Yin,
``Charge redistribution from anomalous magnetovorticity coupling,''
Phys.\ Rev.\ Lett.\  {\bf 117}, 152002 (2016).

\bibitem{Cao:2021jjy}
G.~Cao,
``Macroscopic transports in a rotational system with an electromagnetic field,''
Phys. Rev. D \textbf{104}, 031901 (2021).

\bibitem{Yamamoto:2021gts}
N.~Yamamoto and D.~L.~Yang,
``Helical magnetic effect and the chiral anomaly,''
Phys. Rev. D \textbf{103}, 125003 (2021).

\bibitem{Son:2007ny}
D.~T.~Son and M.~A.~Stephanov,
``Axial anomaly and magnetism of nuclear and quark matter,''
Phys. Rev. D \textbf{77}, 014021 (2008).

\bibitem{Brauner:2016pko}
T.~Brauner and N.~Yamamoto,
``Chiral Soliton Lattice and Charged Pion Condensation in Strong Magnetic Fields,''
JHEP \textbf{04}, 132 (2017).

\bibitem{Cao:2015cka}
G.~Cao and X.~G.~Huang,
``Electromagnetic triangle anomaly and neutral pion condensation in QCD vacuum,''
Phys. Lett. B \textbf{757}, 1-5 (2016).

\bibitem{Wang:2017pje}
L.~Wang and G.~Cao,
``Competition between magnetic catalysis effect and chiral rotation effect,''
Phys. Rev. D \textbf{97}, 034014 (2018).

\bibitem{Wang:2018gmj}
L.~Wang, G.~Cao, X.~G.~Huang and P.~Zhuang,
``Nambu\textendash{}Jona-Lasinio model in a parallel electromagnetic field,''
Phys. Lett. B \textbf{780}, 273-282 (2018).

\bibitem{Cao:2019hku}
G.~Cao,
``The electromagnetic field effects in in-out and in-in formalisms,''
Phys. Lett. B \textbf{806}, 135477 (2020).

\bibitem{Cao:2020pjq}
G.~Cao,
``Effects of a parallel electromagnetic field in the three-flavor Nambu\textendash{}Jona-Lasinio model,''
Phys. Rev. D \textbf{101}, 094027 (2020).

\bibitem{Coleman:1973ci}
S.~R.~Coleman,
``There are no Goldstone bosons in two-dimensions,''
Commun. Math. Phys. \textbf{31}, 259-264 (1973).

\bibitem{Mermin:1966fe}
N.~D.~Mermin and H.~Wagner,
``Absence of ferromagnetism or antiferromagnetism in one-dimensional or two-dimensional isotropic Heisenberg models,''
Phys. Rev. Lett. \textbf{17}, 1133-1136 (1966).

\bibitem{Hohenberg:1967zz}
P.~C.~Hohenberg,
``Existence of Long-Range Order in One and Two Dimensions,''
Phys. Rev. \textbf{158}, 383-386 (1967).

\bibitem{Berezinsky:1970fr}
V.~L.~Berezinsky,
``Destruction of long range order in one-dimensional and two-dimensional systems having a continuous symmetry group. I. Classical systems,''
Sov. Phys. JETP \textbf{32}, 493-500 (1971).

\bibitem{Berezinsky:1972rfj}
V.~L.~Berezinsky,
``Destruction of Long-range Order in One-dimensional and Two-dimensional Systems Possessing a Continuous Symmetry Group. II. Quantum Systems.,''
Sov. Phys. JETP \textbf{34}, 610 (1972).

\bibitem{Kosterlitz:1973xp}
J.~M.~Kosterlitz and D.~J.~Thouless,
``Ordering, metastability and phase transitions in two-dimensional systems,''
J. Phys. C \textbf{6}, 1181-1203 (1973).

\bibitem{Gasser:1983yg}
J.~Gasser and H.~Leutwyler,
``Chiral Perturbation Theory to One Loop,''
Annals Phys. \textbf{158}, 142 (1984).

\bibitem{Wess:1971yu}
  J.~Wess and B.~Zumino,
  ``Consequences of anomalous Ward identities,''
  Phys.\ Lett.\ B {\bf 37}, 95 (1971).

\bibitem{Witten:1983tw}
  E.~Witten,
  ``Global Aspects of Current Algebra,''
  Nucl.\ Phys.\ B {\bf 223}, 422 (1983).
  
\bibitem{Mameda:2014cxa}
K.~Mameda,
``QCD \ensuremath{\theta}-vacua from the chiral limit to the quenched limit,''
Nucl. Phys. B \textbf{889}, 712-726 (2014).

\bibitem{Fetter2003b}
A. L. Fetter and J. D. Walecka, {\it Quantum theory of many-particle systems}, Chapter 13 (McGraw-Hill Book Company, New York, 1971).

\bibitem{Scherer:2002tk}
S.~Scherer,
``Introduction to chiral perturbation theory,''
Adv. Nucl. Phys. \textbf{27}, 277 (2003).

\end{thebibliography}
\end{document}